\documentclass{iopconfser}
\usepackage[compress]{cite}
\usepackage{amsmath, amsthm, amssymb}
\usepackage{amsmath, amsthm, amssymb}
\usepackage{graphicx}
\usepackage{enumitem}
\usepackage{multirow}
\usepackage{framed}

\begin{document}

\title{On defining astronomically meaningful Reference Frames in General Relativity}

\author{L. Filipe O. Costa$^{1}$, Francisco Frutos-Alfaro$^{2}$, José Natário$^{3}$, Michael Soffel$^{4}$}

\affil{$^1$Centro de Matemática (CMAT), University of Minho, Portugal}
\affil{$^2$Space Research Center (CINESPA), University of Costa Rica, Costa Rica}
\affil{$^3$CAMGSD, Instituto Superior Técnico, Portugal}
\affil{$^4$Lohrmann Observatory, Dresden Technical University, Germany}

\email{filipecosta@cmat.uminho.pt}

\begin{abstract}
In a recent paper we 
discussed when it is possible to define reference frames nonrotating with respect to distant inertial reference objects (extension of the IAU reference systems to exact general relativity), and how to construct them. We briefly review the construction, illustrating it with further examples, and 
caution against the recent misuse of
zero angular momentum observers (ZAMOs).
\end{abstract}

\section{Introduction}
The problem of defining astronomically meaningful reference frames is well understood in a post-Newtonian approximation, equipped with coordinate systems (namely the IAU reference system  \cite{Soffel2009IAU,Kopeikin2006RefFrames,KopeikinIAU2009,SoffelBook2009,soffel_Book_ReferenceSystems}) with axes anchored to asymptotically inertial frames, physically materializing in being fixed to distant reference objects (stars or quasars). In the exact theory this is not so well understood, and has been posing difficulties in the interpretation of several solutions \cite{Costa_e_al_Frames}, recently exacerbated in misguided models claiming that relativistic effects can mimic dark matter in explaining the galactic rotation curves.

\section{Reference frames in General Relativity}
To define a reference frame in GR, two fundamental ingredients are needed: (i) a family of observers, i.e., a congruence of timelike curves $\mathcal{O}(u)$, whose 4-velocity $u^{\alpha}$ yields the frame's time-axis \cite{Costa_e_al_Frames,SachsWu1977,Soffel1989book}; (ii) a triad of spatial axes defined along $\mathcal{O}(u)$, see Fig. 1 of \cite{Costa_e_al_Frames}. A coordinate system $\{t,x^{i}\}$, where $\partial_{t} \propto u$, naturally embodies such construction, while additionally providing a means of labeling events.
(Note: Greek letters $\alpha$, $\beta$, $\gamma$, ... denote 4D spacetime indices, running 0-3, and Roman letters $i,j,k,...$ are spatial indices.)

\subsection{Shearfree frames \label{Shearfreeframes}}
\begin{figure}
\includegraphics[width=0.95\columnwidth]{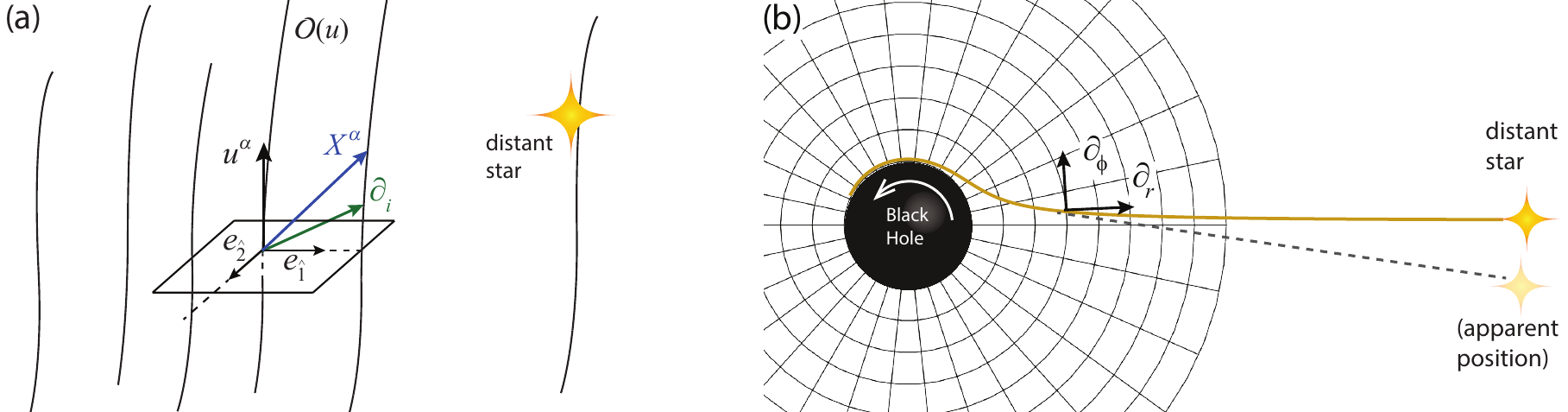} \caption{\label{fig:ShearfreeCongruence} An astronomically meaningful reference frame: coordinate system adapted to a \textit{shearfree} congruence of observers $\mathcal{O}(u)$ with asymptotically vanishing acceleration and vorticity. (a) Connecting vectors ($X^{\alpha}$ and $\partial_{i}$) have fixed direction in the orthonormal triad $\{{\bf e}_{\hat{\imath}}\}$, yielding a grid of points everywhere at fixed directions with respect to each other, and anchored to inertial frames at infinity.  (b) 3D space representation of such a grid. In the special case of conformally stationary spacetimes (e.g. the Kerr metric) this materializes in that light rays from remote sources arrive at fixed directions in such a frame.}
\end{figure}

\begin{figure}
\begin{tabular}{lll}
{\footnotesize{}\raisebox{5ex}{}}\raisebox{1ex}{(a)~Rigidly
rotating congruence} & \raisebox{1ex}{(b)~Purely expanding congruence} & \raisebox{1ex}{(c)~Shearing congruence}\tabularnewline
\includegraphics[width=0.33\textwidth]{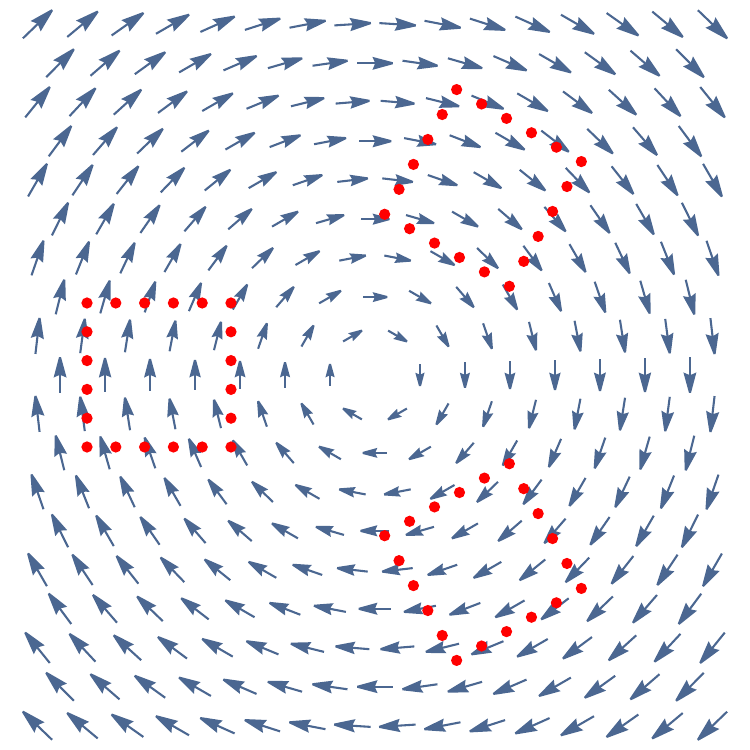} & \includegraphics[width=0.33\textwidth]{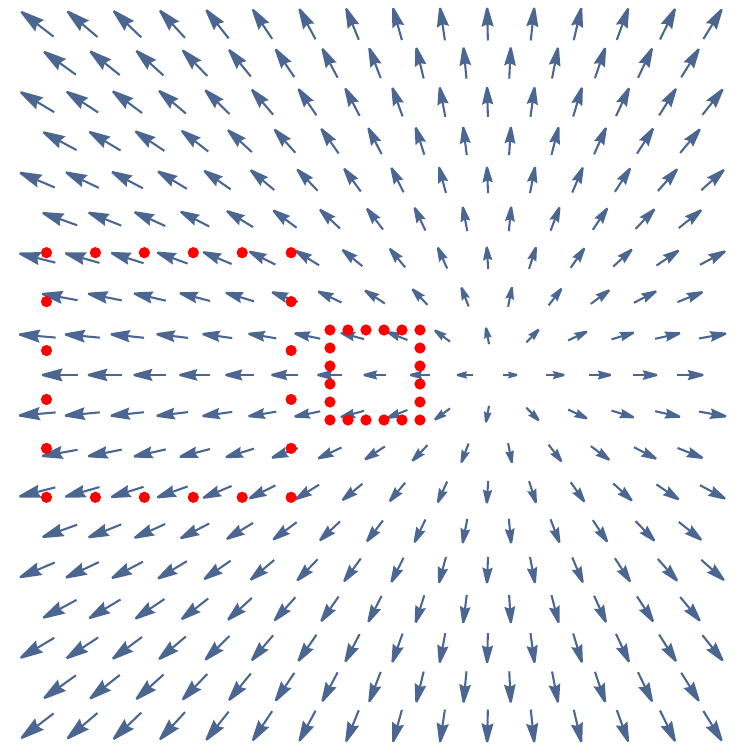} & \includegraphics[width=0.33\textwidth]{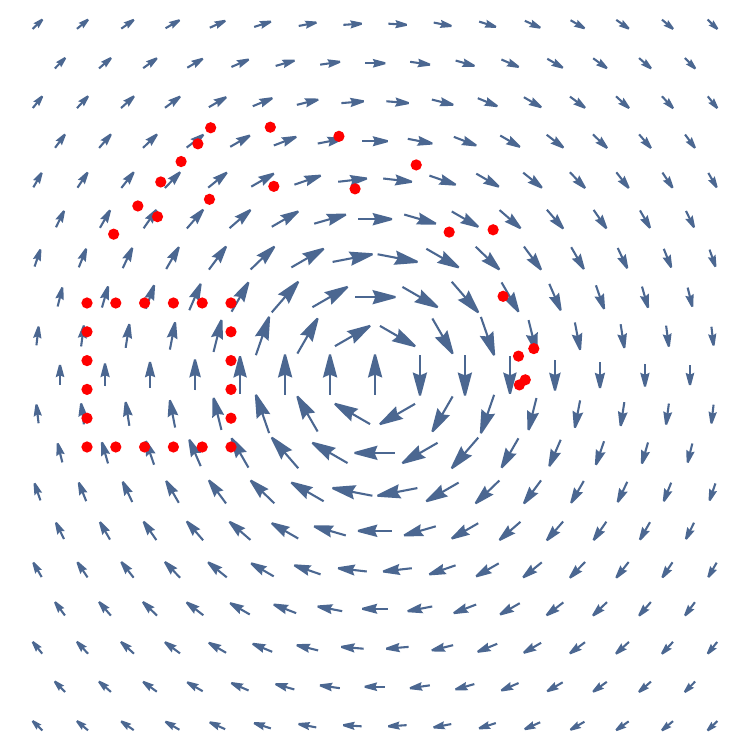}\tabularnewline
{\small{}$u^{\alpha}=\gamma(\partial_{t}+\Omega\partial_{\phi})\ ;$}{\footnotesize{}$\quad$}{\small{}$\Omega={\rm constant}$} & {\small{}$u^{\alpha}=\zeta(t\partial_{t}+r\partial_{r})\ ;$}{\footnotesize{}$\quad$$\zeta=1/\sqrt{t^{2}-r^{2}}$} & {\small{}$u^{\alpha}=\gamma(\partial_{t}+\Omega\partial_{\phi})\ ;$}{\footnotesize{}$\quad$}{\small{}$\Omega={\rm const.}/r^{2}$ }\tabularnewline
\end{tabular} \caption{\label{fig:CongruencesFlat} Examples of congruences in flat spacetime: (a) rigidly rotating congruence; (b) congruence associated to Milne's universe \cite{GriffithsPodolsky2009} (purely expanding: $a^{\alpha}=\omega^{\alpha}=\sigma_{\mu \nu}=0$); (c) free or irrotational vortex (e.g., a whirlpool)---shearing congruence. The plots depict the spatial velocities $dx^{i}/dt=u^{i}/u^{0}$. Red dots show the evolution of an initial squared array of observers: in (a) and (b) the squared shape is preserved along the congruence, reflecting the preservation of angles between observers. In (c), by contrast, the shape is completely distorted as it evolves along the congruence.}
\end{figure}

The shearfree property is paramount in this context. Consider vectors connecting the worldlines of neighboring observers, defined by the Lie transport condition $\mathcal{L}_{{\bf u}}X^{\hat{\alpha}}=0$ (corresponding to vectors connecting events with the same proper-time separation on adjacent worldlines). Consider also an orthonormal frame `adapted to the observers', $\{{\bf u},{\bf e}_{\hat{\imath}}\}$; in such frame, $\mathcal{L}_{{\bf u}}X^{\hat{\alpha}}=0$ yields \cite{Costa_e_al_Frames,Analogies}
\begin{equation}
\dot{X}_{\hat{\imath}}=\frac{1}{3}\theta X_{\hat{\imath}}+\sigma_{\hat{\imath}\hat{\jmath}}X^{\hat{\jmath}}+\epsilon_{\hat{\imath}\hat{k}\hat{\jmath}}\left(\omega^{\hat{k}}-\Omega^{\hat{k}}\right)X^{\hat{\jmath}},\label{eq:ConnectingVector}
\end{equation}
where $\omega^{\alpha}=\epsilon^{\alpha\beta\gamma\delta}u_{\gamma;\beta}u_{\delta}/2$, $\sigma_{\alpha\beta}=h_{\alpha}^{\mu}h_{\beta}^{\nu}u_{(\mu;\nu)}-\theta h_{\alpha\beta}/3$ , and $\theta=u_{\ ;\alpha}^{\alpha}$  are, respectively, the vorticity, shear, and expansion of the observer congruence, and $h_{\ \beta}^{\alpha}\equiv\delta_{\ \beta}^{\alpha}+u^{\alpha}u_{\beta}$ the space projector orthogonal to $u^{\alpha}$. The quantity $\Omega^{\alpha}$ is the angular velocity of rotation
of the spatial triad ${\bf e}_{\hat{\imath}}$ with respect to Fermi-Walker transport: $(\nabla_{\mathbf{u}}\mathbf{e}_{\hat{\imath}})^{\hat{\alpha}}-u^{\hat{\alpha}}a_{\hat{\imath}}=\epsilon_{\ \hat{\imath}\hat{\nu}\hat{\mu}}^{\hat{\alpha}}\Omega^{\hat{\mu}}u^{\hat{\nu}}$, which is not fixed by the congruence. Choosing $\Omega^{\alpha}=\omega^{\alpha}$---the natural choice, corresponding to axes co-rotating, or  `adapted' to the congruence  \cite{MassaZordan,MassaII,ManyFaces, Analogies}---Eq. \eqref{eq:ConnectingVector} tells us that, if the shear vanishes ($\sigma_{\alpha\beta}=0$), then $\dot{X}^{\hat{\imath}}=\frac{1}{3}\theta X^{\hat{\imath}}$, i.e., the connecting vector's direction is fixed in the triad $\{e_{\hat{\imath}}\}$. Since the triad is orthonormal, this means that neighboring \textit{observers remain at fixed angles with respect to each other}, see Fig. \ref{fig:ShearfreeCongruence}(a). The congruence represents in this case the history of a grid of points \textit{everywhere} at fixed directions 
[Fig. \ref{fig:ShearfreeCongruence}(b)]; measuring rotations with respect to such a grid amounts to measuring it with respect to distant observers/objects. 
This is exemplified in Figs. \ref{fig:CongruencesFlat}(a) and (b), respectively, with a rigidly rotating and an expanding congruence in flat spacetime. 
In Fig. \ref{fig:CongruencesFlat}(c) we consider a shearing congruence, showing this construction to break down.

\subsection{Metric tensor of spacetimes admitting shearfree observer congruences \label{Secmetric}}
The line element of an arbitrary spacetime can be written as 
\begin{equation}
ds^{2}=-e^{2\Phi(t,x^{k})}[dt-\mathcal{A}_{i}(t,x^{k})dx^{i}]^{2}+h_{ij}(t,x^{k})dx^{i}dx^{j}\label{eq:GenMetric}
\end{equation}
where $h_{ij}=g_{ij}+e^{2\Phi}\mathcal{A}_{i}\mathcal{A}_{j}$ equals the space components of the projector $h_{\alpha \beta}$. The shearfree condition $\sigma_{\alpha\beta}=0$ is equivalent to $\mathcal{L}_{{\bf u}}h_{\alpha\beta}=2\theta h_{\alpha\beta}/3$, which amounts to $h_{\alpha\beta}=f\chi_{\alpha\beta}$, with $\mathcal{L}_{{\bf u}}\chi_{\alpha\beta}=0$ and $f$ a solution of the equation $\mathcal{L}_{{\bf u}}f-2\theta f/3=0$, which is the condition of ``conformal rigidity'' (Bel-Llosa, 1995). 
In a coordinate system adapted to the observers (${\bf u} \propto \partial_{t}$), this yields $\partial_{t}\chi_{\alpha\beta}=0\Rightarrow h_{\alpha\beta}=f(t,x^{i})\chi_{\alpha\beta}(x^{i})$. By \eqref{eq:GenMetric} it follows that the existence of shearfree observer congruences in a given spacetime is equivalent to it admitting a coordinate system where the metric takes the form \eqref{eq:GenMetric} with
\begin{equation}
h_{ij}(t,x^{k})=f(t,x^{k})\chi_{ij}(x^{k})\ .\label{eq:shear-freeMetric}
\end{equation}
Notice the restrictive condition: whereas a general metric possesses 6 ``free'' functions of 4 variables after gauge fixing (made explicit in the synchronous gauge, e.g. Eq. (9) of \cite{Costa_e_al_Frames}), the metric \eqref{eq:shear-freeMetric} has only 5 functions ($\Phi$, $\mathcal{A}_{i}$,$f$) of 4 variables, plus 6 functions ($\chi_{ij}$) of 3 variables (recall that a function of 4 variables amounts to infinitely many functions of 3 variables).

\subsection{Astronomically meaningful reference frames \label{astroframes}}
The axes of astronomical reference frames are set fixed with respect to distant inertial reference objects (stars or quasars) 
whose proper motions are negligible, thereby defining the directions of idealized inertial frames at infinity.
If the vorticity of a shearfree congruence asymptotically vanishes, $\lim_{r\rightarrow\infty}\omega^{\alpha}=0$, it follows from Sec. \ref{Shearfreeframes} that it represents a grid of points rotationally anchored to inertial frames at infinity. If moreover the acceleration $a^{\alpha}\equiv u^{\beta}u_{\ ;\beta}^{\alpha}$ asymptotically vanishes, $\lim_{r\rightarrow\infty}a^{\alpha}=0$, then it can be said to be anchored to distant stars or quasars, see Fig. \ref{fig:ShearfreeCongruence}. We thus arrive at the following: 
\begin{framed}
If a spacetime admits a non-shearing congruence of observers which, at infinity, has zero
vorticity and acceleration, then a coordinate system \eqref{eq:shear-freeMetric} where they are at rest has axes $\partial_{i}$ locked to the asymptotic rest frame of the distant quasars, generalizing the IAU reference system to exact GR.
\end{framed}
Astronomical reference frames are physically set up aiming telescopes at the reference stars or quasars. 
Light emitted from distant sources arrives at fixed spatial directions in the basis $\{\partial_{i}\}$ (or $\{e_{\hat{\imath}}\}$) in the case of conformally stationary spacetimes, whose metric has the form $ds^{2}=\psi(t,x^{i})\Psi_{\alpha\beta}(x^{i}) dx^{\alpha}dx^{\beta}$---even more restrictive than \eqref{eq:shear-freeMetric} \cite{Costa_e_al_Frames}. Otherwise, time-dependent gravitational lensing effects cause the spatial axes 
to only approximately coincide with the direction of light rays received from distant objects.

\section{Stationary spacetimes and zero angular momentum observers (ZAMOs)}
Stationary spacetimes are characterized by admitting timelike Killing vector fields; their integral lines are rigid observer congruences ($\sigma_{\alpha \beta}=\theta=0$) and, in coordinate systems where they are at rest, the metric takes the form \eqref{eq:GenMetric}-\eqref{eq:shear-freeMetric} with $f=1$ and $\Phi$ and $\mathcal{A}$ time-independent. If such a congruence exists which is moreover asymptotically inertial ($\lim_{r\rightarrow\infty}\omega^{\alpha}=\lim_{r\rightarrow\infty}a^{\alpha}=0$), it forms a \textit{rigid} grid anchored to inertial frames at infinity, and the associated coordinate system is the generalized astronomical frame in such setting. Examples include the Boyer-Lindquist coordinates in black hole spacetimes and the star-fixed coordinates for the van Stockum rotating cylinder \cite{Cilindros,Stockum1938,Costa_e_al_Frames}.  
 
In some recent literature \cite{BG,Crosta2018,RuggieroBG,Beordo2024,CooperstockTieu,CarrickCooperstock}, however, a different class of observers---the ZAMOs---is confused with observers at rest in astronomical frames, and used to compute rotation curves in purported galactic toy models. ZAMOs are defined in axistationary spacetimes as observers of the form $u_{{\rm Z}}^{\alpha}=u_{{\rm Z}}^{0}(\delta_{0}^{\alpha}+\Omega_{{\rm ZAMO}}\delta_{\phi}^{\alpha})$ for which the angular momentum vanishes, $(u_{{\rm Z}})_{\phi}=0$. However, when frame dragging is present ($g_{0i}\ne 0$ in the coordinate system defined in Sec. \ref{astroframes}),
\begin{itemize}[itemsep=0pt, parsep=0pt, topsep=0pt, partopsep=0pt]
\item these observers move circularly with angular velocity  $\Omega_{{\rm Z}}\equiv u_{{\rm Z}}^{\phi}/{u_{{\rm Z}}^{0}}=-g_{0i}/{g_{00}}$ relative to the coordinate system rigidly fixed to inertial frames at infinity (i.e., to the actual astronomical frame).
\item they are a shearing congruence: $\sigma_{{\rm Z}}^{\alpha\beta}=u_{{\rm Z}}^{0}\Omega_{{\rm Z}}^{,(\alpha}\delta_{\phi}^{\beta)}\ne0$, since $\Omega_{{\rm Z}}$ is not constant. 
\end{itemize}
Such confusion has grave consequences: in the Kerr spacetime, one would conclude that the black hole does not rotate, since the ZAMOs at the horizon comove with it. In the alleged galactic model \cite{BG,Crosta2018,Beordo2024}, the reported flat rotation curves (implying differential rotation) are an artifact of the ZAMOs’ circular motion and shear, since the model is rigid and actually static \cite{Costa_e_al_Frames} relative to asymptotic inertial frames.

Such misunderstandings seem to stem from a terminology confusion in the literature, where similar names mean different things: ZAMOs are said to be locally nonrotating with respect to the local geometry  \cite{Misner:1974qy,Bardeen1970ApJ,BardeenPressTeukolsky} (in the sense of measuring no Sagnac effect), a nonrotating congruence \cite{StephaniExact,WyllemanBeke2010} [in the sense of having no vorticity; like the irrotational vortex in Fig. \ref{fig:CongruencesFlat} (c)], and tetrads carried by them ``locally nonrotating frames" \cite{Bardeen1970ApJ,BardeenPressTeukolsky}; see \cite{Costa_e_al_Frames} Sec. III.E.2 for details. It is crucial (due to frame-dragging) to not confuse any of these notions with the ``kinematically nonrotating local reference system'' used in astrometry \cite{KlionerSoffel_Nonrotating1998,Klioner_Microarcsecond_2003}, which is a local system of axes nonrotating with respect to distant reference objects. 

\section{Conclusions}
We showed that an extension of the IAU reference system to the exact theory, preserving the property of defining fixed directions with respect to distant
reference inertial objects, is possible in spacetimes admitting shearfree observer congruences which are also asymptotically vorticity and acceleration-free, and obtained the general form of the metric in the coordinates adapted to such observers.
Examples of such spacetimes include stationary asymptotically flat (e.g. Kerr) and some non-asymptotically flat spacetimes (e.g. NUT, van Stockum-Weyl class, cosmic strings \cite{Costa_e_al_Frames}), and shearfree cosmological models  \cite{KrazinskiShearfree,BarnesShearfree,Sussman_1993} (e.g. FLRW).
The construction allows also to avoid inappropriate choices of reference observers, such as the ZAMOs at the origin of misguided galactic models in recent literature. 

\section*{Acknowledgments}
LFC acknowledges the support of CMAT (Centro de Matemática da Universidade do Minho) through FCT/Portugal projects UID/00013:Centro de Matemática da Universidade do Minho (CMAT/UM) and UID/00013/2025. JN was partially supported by FCT/Portugal through CAMGSD, IST-ID (projects UIDB/04459/2020 and UIDP/04459/2020) and project 2024.04456.CERN, and also by the H2020-MSCA-2022-SE project EinsteinWaves, GA no. 101131233.

\bibliography{Ref}

\end{document}